\documentclass[aps,prl,twocolumn,showpacs,preprintnumbers,amsmath,amssymb]{revtex4}
\usepackage{graphicx}
\usepackage{natbib}
\usepackage{dcolumn}
\usepackage{bm}

\begin{document}

\title{Effect of magnetic state on the $\gamma -\alpha$ transition in iron:
First-principle calculations of the Bain transformation path}

\author{S. V. Okatov}
\affiliation{Institute of Quantum Materials Science, Ekaterinburg
620107, Russia}
\author{A.~R. Kuznetsov, Yu.~N. Gornostyrev}
\affiliation{Institute of Metal Physics, Russian Academy of
Sciences, Ural Division, Ekaterinburg 620041, Russia}
\affiliation{Institute of Quantum Materials Science, Ekaterinburg
620107, Russia}
\author{V. N. Urtsev}
\affiliation{Research and Technological Center "Ausferr",
Magnitogorsk, 455023, Russia}
\author{M. I. Katsnelson}
\affiliation{Institute for Molecules and Materials, Radboud
University Nijmegen, NL-6525 AJ Nijmegen, the Netherlands}

\date{\today}


\begin{abstract}
Energetics of the fcc ($\gamma$) - bcc ($\alpha$) lattice
transformation by the Bain tetragonal deformation is calculated
for both magnetically ordered and paramagnetic (disordered local
moment) states of iron. The first-principle computational results
manifest a relevance of the magnetic order in a scenario of the
$\gamma$ - $\alpha$ transition and reveal a special role of the
Curie temperature of $\alpha$-Fe, $T_C$, where a character of the
transformation is changed. At a cooling down to the temperatures
$T < T_C$ one can expect that the transformation is developed as a
lattice instability whereas for $T > T_C$ it follows a standard
mechanism of creation and growth of an embryo of the new phase. It
explains a closeness of $T_C$ to the temperature of start of the
martensitic transformation, $M_s$.
\end{abstract}

\maketitle


Deeper understanding of mechanisms of polymorphous $\gamma$ -
$\alpha$ transformation in iron and its alloys is of fundamental
importance for both metallurgical technologies \cite{1,2} and for
a general theory of phase transitions in solids
\cite{3,4,olson06}. Despite numerous investigations an issue of a
mechanism of a new phase nucleation in the course of $\gamma$ -
$\alpha$ transformation remains open (see, e.g., discussion in
Ref. \onlinecite{4}).

It is well known \cite{1} that a character and rate of the
transformation in iron and iron-based alloys is changed
drastically below some temperature $M_s$, namely, at $T < M_s$ it
occurs by a fast cooperative shear deformation of atoms
(martensitic mechanism) whereas for higher temperatures (but lower
than the temperature of phase equilibrium $T_{\gamma - \alpha}$)
the transformation develop much slower and a formation and growth
of grains of $\alpha$-phase is observed. In pure iron the
temperature of start of the martensitic transformation $M_s
\approx 1020K$ \cite{5,6} which is only 30 K lower than the Curie
temperature of $\alpha$-Fe, $T_C$, in accordance with an old idea
of Zener \cite{7} suggesting that $M_s \approx T_C$. It is know
also that in the system Fe-C a process of the $\gamma$ - $\alpha$
transformation becomes much faster at a cooling down below $T_C$
\cite{8}. It is commonly accepted now that magnetic degrees of
freedom play a crucial role in the phase equilibrium in iron
\cite{9,10,11,danil}. However, a mechanism of their effect on
kinetics of the transformation remains unclear.

Several ways to transform the crystal lattice from fcc ($\gamma$)
to bcc ($\alpha$) structures have been suggested, among them
schemes of Bain \cite{12} (a tetragonal deformation along the
$<001>$ axis) and of Kurdumov-Zaks \cite{1} (two shear
deformations) are the best known. To determine the deformation
path and estimate energy barriers calculations of total energy in
a configurational space of lattice deformations are required. For
the Kurdumov-Zaks path this is a rather cumbersome and
computationally expensive problem. Therefore first-principle
calculations of energetics of polymorphous transformation in iron
were carried out, up to now, only for the Bain transformation path
\cite{11,13,14,15,16,17,18}. It turned out that energetics of the
transformation is essentially dependent on magnetic structure of
iron. In particular, ferromagnetic (FM) state of the fcc iron is
unstable with respect to tetragonal deformation, with energy
minima at $c/a$ ratio equal 1 (bcc) or $c/a \approx 1.5$ (fct
structure) \cite{11,13,14}. Antiferromagnetic (AFM) state of the
fcc iron has lower energy than the FM one and show a monotonous
increase of energy along $\gamma \rightarrow \alpha$
transformation path \cite{13,14}.

Since the magnetic ground state is different for $\alpha$- and
$\gamma$-Fe, an investigation of energetics of the transformation
path for a given collinear magnetic structure
\cite{11,13,14,15,16,17} is not enough to describe properly the
$\alpha - \gamma$ transition. As was shown in Refs.
\onlinecite{19,20,21} the true magnetic ground state of fcc Fe is
rather complicated and, in general, noncollinear (depending on the
lattice constant). Evolution of magnetic state of Fe along the
Bain path was studied in Ref. \onlinecite{18} by the TB-LMTO
method in atomic sphere approximation (ASA) \cite{22}. It was
demonstrated that, for an essential part of the Bain path,
noncollinear magnetic structures take place which are replaced by
FM ordering for $c/a$ smaller than some critical value.

The calculations \cite{13,14,15,17} have been carried out for a
fixed value of volume per atom or lattice constant $a$ that
corresponds to so-called epitaxial Bain path describing the
transformation for iron films on a substrate. The latter leads to
some restrictions on geometry of the transformation which takes
place at low temperatures in magnetically ordered state. The
results of Ref. \onlinecite{18} allow us to predict a type of
magnetic structures which can be realized in the epitaxial iron
films. At the same time, the $\gamma - \alpha$ transformation in
the bulk occurs at high temperatures when the magnetic state is
disordered. To describe this situation, calculations for
paramagnetic iron are required.

In this work we analyse in detail the Bain deformation path for
both noncollinear and paramagnetic states using the methods of
spin spirals (SS) and disordered local moments (DLM),
respectively. As a result, we clarify the reasons for essential
differences in mechanisms of the polymorphous transformation below
and above the Curie temperature $T_C$.

The calculations have been carried out using the VASP (Vienna
Ab-initio Simulation Package) \cite{23,24,25} with first-principle
pseudopotentials constructed by the projected augmented waves
(PAW) \cite{26}, and the generalized gradient approximation (GGA)
for the density functional in a form \cite{27} with the
parametrization \cite{28} (see also comments in Ref.
\onlinecite{31} concerning use of the GGA for noncollinear case).
The PAW potential with energy mesh cutoff 530 eV, and uniform
$k$-point 12$\times$12$\times$12 mesh in the Monkhorst-Park scheme
\cite{MP} were used.

To describe the noncollinear magnetic state the model of flat
SS \cite{31} has been used with the magnetization rotation
around wave vector of SS, ${\bf q}$ which chosen along the axis of the
Bain tetragonal deformation $<001>$.
The values $q = 0$ and $q = 0.5$ (in units of $2\pi /c$)
correspond to the FM and AFM states, respectively. To calculate
the energy of Fe with SS magnetic structure we employed the PAW
formalism described in Ref. \onlinecite{hobbs} and implemented in
the VASP code. We optimized the volume per atom for given values
of $q$ and tetragonal deformation.

Paramagnetic state of iron was modeled by the disordered local
moment (DLM) method \cite{30}. To this aim, we used 27-atom
supercell (Fig. 1) with a given random distribution of magnetic
moments (with zero total magnetic moment) which was kept fixed at
the self-consistency process. The latter was provided by using
constrained density functional approach \cite{constrain}. As well
as for the SS case we optimized the volume per atom for a given
tetragonal deformation.

\begin{figure}[!thb]
\centerline{
\includegraphics[width=8.10cm,clip,angle=0]{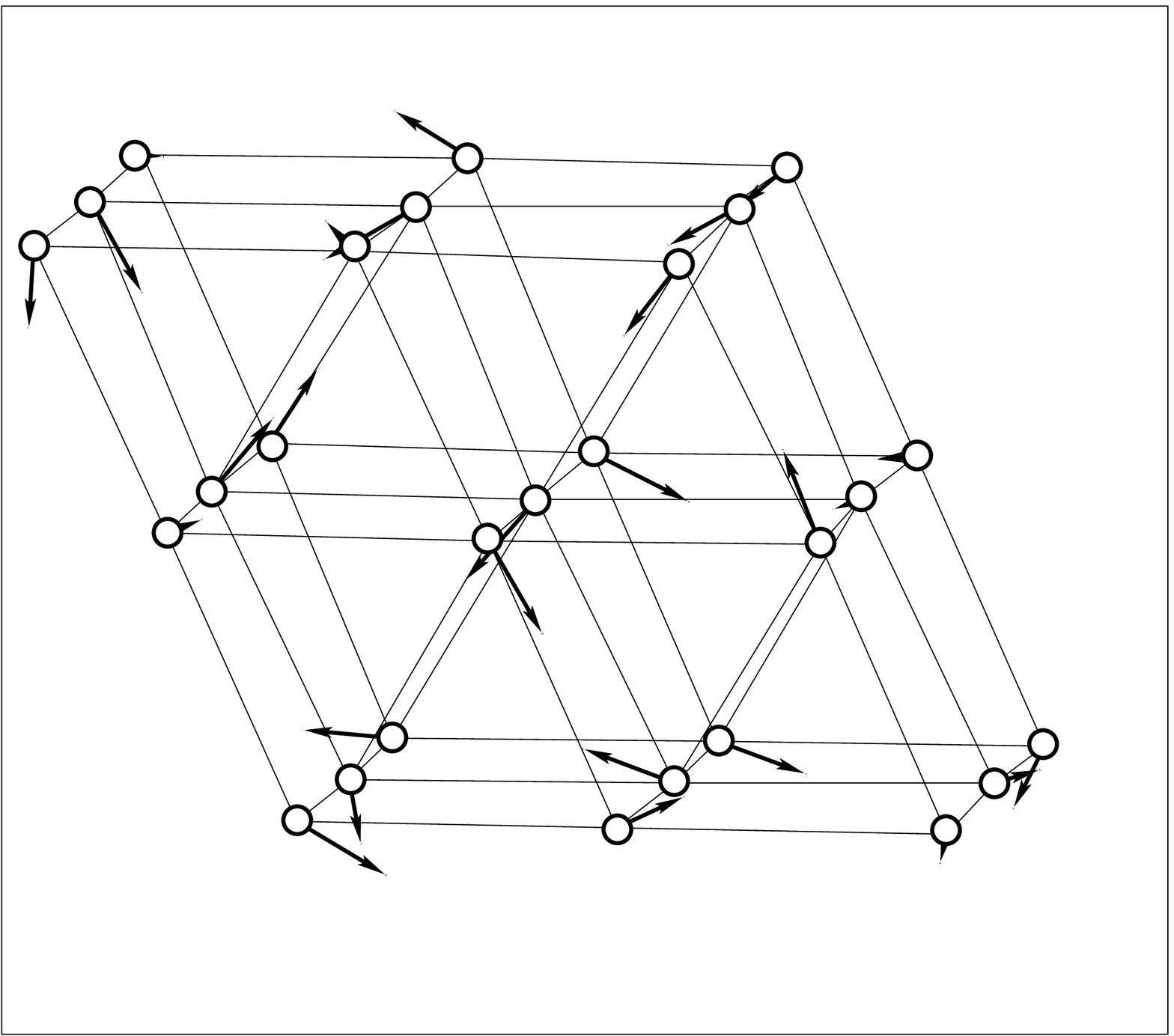}}
\vspace{-0.0cm} \caption[a] {{\small Supecell used in the DLM
calculation. Arrows show the magnetic moments.
 }
}
\label{fig:1}
\end{figure}

\begin{figure}[!thb]
\centerline{
\includegraphics[width=9.00cm,clip,angle=0]{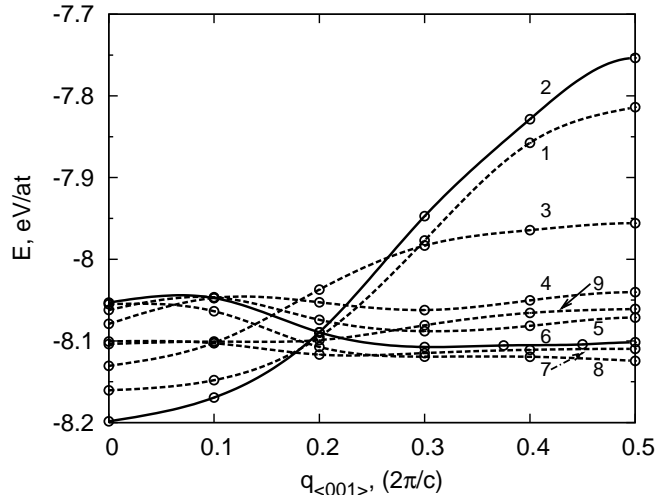}}
\vspace{-0.0cm}
\caption[a] {{\small
Dependence of the total energy per atom for SS on the
wave vector ${\bf q} \parallel <001>$ for different values of
tetragonal deformation. Curves 1 to 9 correspond to $c/a = 0.9;
1.0; 1.2; 1.3; 1.35; \sqrt{2}; 1.5; 1.6; 1.7$, respectively; $c/a
=1$ for bcc and $c/a = \sqrt{2}$ for fcc structures.
}
}
\label{fig:2}
\end{figure}

Computational results for SS are shown in Fig. 2. One can see that
the minimum of the total energy is reached for bcc FM (curve 2)
and fct AFM (curve 8), with a transition between these two states
via noncollinear magnetic structure with the wave vectors $q
\simeq 0.25 - 0.35$. The energy difference between the FM and AFM
states is small for $1.3 < c/a < 1.5$ and increases, almost in
order of magnitude, for $c/a <1.3$. This agrees qualitatively with
the previous results \cite{18} of the TB-LMTO-ASA calculations
that the tetragonal deformation induces a magnetic transition.

Similar to Refs. \onlinecite{18,29,31} we found that the energy
minimum for the fcc Fe corresponds to $q \simeq 0.3$ (curve 6 in
Fig. 2). Optimization of the volume makes the $q$-dependence of
the total energy rather flat, thus, the energy at $q = 0.3$ is
only 6 meV/atom lower than for the AFM fcc state. The FM fcc
structure ($q = 0$) has the energy in 54 meV/atom higher than for
$q = 0.3$. It is characterized by much larger values of magnetic
moment and volume per atom (11.86 ~\AA$^3$ and 2.3 $\mu_B$ at
$q=0$ vs 10.74 ~\AA$^3$ and 1.4 $\mu_B$ at $q=0.3$). One can see
in Fig. 2 that the FM state of fcc iron is unstable with respect
to the tetragonal deformation; if one keeps the FM order the fcc
iron will reconstruct spontaneously to bcc or fct phase, without
essential changes of volume and magnetic moment per atom.

Fig. 3 presents the total energy along the Bain path for optimized
(with a given $c/a$ ratio) SS magnetic structure, together with the
results for the collinear (FM and AFM) case, for the
paramagnetic case (DLM) and for partially disordered magnetic structure
(DLM$_{0.5}$). The results for the FM and AFM magnetic
structures are close to those obtained earlier by full-potential
FLAPW method \cite{13}. Taking into account noncollinear magnetic
configurations allows us to describe a continuous transition
between FM and AFM states and to estimate the energy barrier
$E_b$ resulting from the Bain deformation. We have found $E_b
\simeq 20$ meV/atom for fcc-bcc and $E_b \simeq 40$ meV/atom for
fct-bcc transitions. The corresponding values found in Ref.
\onlinecite{13} are 48 meV/atom and 63 meV/atom, respectively which
is essentially larger than our data. The difference is mainly due
to optimization of volume per atom used in our calculations (in
Ref. \onlinecite{13} fixed values of the volume were used).

\begin{figure}[!thb]
\centerline{
\includegraphics[width=9.00cm,clip,angle=0]{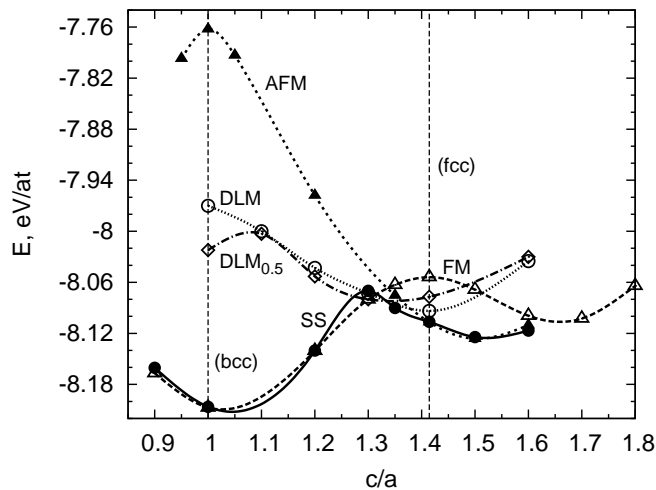}}
\vspace{-0.0cm} \caption[a] {{\small Variation of the total energy
per atom along the Bain deformation path for different magnetic
states. FM (empty triangles) and AFM (solid triangles) label
collinear ferromagnetic and antiferromagnetic structures, SS
(solid circles) - spin-spiral state, DLM (empty circles) -
disordered local moments, DLM$_{0.5}$ (empty diamonds) - DLM state
with the total magnetic moment equal to half of that for the FM
state.
 }
}
\label{fig:3}
\end{figure}

Comparing Figs. 2 and 3 one can see that the barrier position
corresponds to the value of tetragonal deformation where magnetic
state switches from SS ($q \approx 0.3$) to FM ($q = 0$).
Interestingly, the critical value of the Bain deformation found
for the SS state turns out to be close to the crossing point of
the corresponding curves for FM and AFM configurations.

For the paramagnetic (DLM) state the energy has a minimum
corresponding to the fcc structure and is close to the energy of
SS and AFM states for $1.3 < c/a < \sqrt{2}$ (Fig. 3). For $c/a <
1.3$ the energy grows monotonously with the $c/a$ decrease
reaching the maximum for the bcc state ($c/a = 1$). This maximum
is much higher than the energy for FM bcc and DLM fcc
configurations (by 220 and 110 meV/atom, respectively). Thus,
reconstruction of crystal lattice of Fe from fcc to bcc by the
Bain deformation without change of magnetic configuration does not
lead to any energy gain. To model the state of iron at finite
temperatures below $T_C$ we have performed also calculations for
the DLM state with total magnetization $M = 0.5 M_{max}$ where
$M_{max}$ is the magnetization in FM state for given $c/a$ ratio.
The results are not essentially different from the paramagnetic
case except for region close to bcc structure (compare the curves
DLM and DLM$_{0.5}$ in Fig. 3). So, the energetics of the Bain
transformation path changes drastically when the magnetic state
becomes close to the ferromagnetic.

It is not surprising, of course, that the energy of bcc Fe is much
higher in the DLM state than in ferromagnetic one, in light of a
common opinion (originated from a seminal work by Zener \cite{33})
that it is ferromagnetism that stabilizes $\alpha$-Fe. According
to Ref. \onlinecite{32} a strong enough short-range magnetic order
survives in iron above $T_C = 1043$ K which is probably essential
to explain a stability of bcc phase in some temperature region
above $T_C$ \cite{10,33}. To consider this problem quantitatively
one has to be able to calculate phonon and magnetic contributions
to entropy of different phases which is rather complicated.
However, since $T_C$ is much higher than typical phonon energies
the lattice vibration entropy in both phases are approximately
constant in the temperature region under consideration and
therefore the phonon contribution to the difference of free
energies between $\alpha$- and $\gamma$-phases is a smooth
function linearly dependent on the temperature. At the same time,
magnetic contribution to the total energy is strongly dependent on
the magnetic configuration and thus on the temperature, as follows
from the presented results.

When cooling down iron in a temperature interval $T_C < T <
T_{\gamma - \alpha}$, the bcc structure becomes preferable due to
entropy contribution in free energy. Since the difference between
free energies of the bcc and fcc phases is zero at $T_{\gamma -
\alpha}$ and changes slowly with the temperature, the moving force
of the phase transition is relatively weak. As a result,  the
$\gamma - \alpha$ transformation develops in this situation by the
classical nucleation and growth mechanism.

When decreasing the temperature below $T_C$ the FM state of bcc Fe
arises which has the energy {\it much} lower than that of
paramagnetic fcc state (by 110 meV/atom). At the same time,
lattice reconstruction from paramagnetic fcc to FM bcc state
requires to overcome a rather low barrier which height ($\sim 20$
meV/atom, ot $\sim 250$ K/atom) is small in comparison with the
temperature $T \sim M_s$. Thus one can expect that cooling down to
$T < T_C$ will initiate martensitic mechanism of $\gamma - \alpha$
transformation, via a development of lattice instability
\cite{34}. This conclusion differs essentially from that of Ref.
\onlinecite{11} where use the less accurate LMTO method with model
Stoner parameters has resulted in three times larger barrier
height than in our calculations.

Thus, $T_C$ of $\alpha$-Fe plays a role of a special point where
kinetics of the transformation is changed, due to a crucial role
of magnetic degrees of freedom in energetics of iron which is a
main conclusion from our computational results.


The work was supported by the Stichting Fundamenteel Onderzoek der
Materie (FOM), The Netherlands, and by Russian Basic Research
Foundation (Grant 06-02-16557).

\end{document}